\documentclass[aps,prd,amsmath,showpacs,showkeys,preprint]{revtex4}
\usepackage{amsmath}
\usepackage[cp1251,koi8-r]{inputenc}
\usepackage[english]{babel}
\usepackage{graphics}
\usepackage[dvips]{color}
\usepackage{epsfig}
\newcommand{\be}{\begin{equation}}
\newcommand{\ee}{\end{equation}}
\newcommand{\bn}{\begin{eqnarray}}
\newcommand{\en}{\end{eqnarray}}
\newcommand{\bd}{\begin{displaymath}}
\newcommand{\ed}{\end{displaymath}}
\newcommand{\bnn}{\begin{eqnarray*}}
\newcommand{\enn}{\end{eqnarray*}}
\def\Ref#1{(\ref{#1})}
\def\Journal#1#2#3#4#5#6{#1, \ #2, \ #3 \  #4 \ (#5) \ #6.}

\begin{document}
\inputencoding{cp1251}
\title{Age-related alterations of relaxation processes and non-Markov effects in
stochastic dynamics of R-R intervals variability from human ECGs}
\author{Renat M. Yulmetyev} \email{rmy@theory.kazan-spu.ru; rmy@dtp.ksu.ras.ru}
\author{Sergey A. Demin} \author{Oleg Yu. Panischev}
\affiliation{Department of Physics, Kazan State Pedagogical
University, 420021 Kazan, Mezhlauk Street, 1 Russia}
\author{Peter H\"anggi}\affiliation{Department of
Physics, University of Augsburg, Universit\"atsstrasse 1, D-86135
Augsburg, Germany}
\begin{abstract}

In this paper we consider the age-related alterations of heart
rate variability on the basis of the study of non-Markovian
effects. The age dynamics of relaxation processes is
quantitatively described by means of local relaxation parameters,
calculated by the specific localization procedure. We offer a
quantitative informational measure of non-Markovity to evaluate
the change of statistical effects of memory. Local relaxation
parameters for young and elderly people differ by 3.3  times, and
quantitative measures of non-Markovity differ by 4.2  times. The
comparison of quantitative parameters allows to draw conclusions
about the reduction of relaxation rate with ageing and the higher
degree of the Markovity of heart rate variability of elderly
people.
\end{abstract}
\pacs{05.40.Ca; 05.45.Tp; 87.19.Hh; 87.75.-k}

\keywords{Discrete non-Markov processes; Time-series analysis;
Heart rate; Relaxation processes; Complex systems}
\maketitle

\section{Introduction}
The ageing of a human organism has been in the focus of attention
in physics of live systems over the past years. One of the most
significant systems of vital activity of a human body is its
cardiovascular system. Today there are a number of scientific
studies on the problems of biological ageing of the cardiovascular
human system. The latter is extremely sensitive to age-related as
well as pathological changes in a human organism. Thus not only
physiologists, biologists and physicians have been involved in
this studies  but also experts from other natural-science areas.
Conditions of a human heart are estimated by means of various
parameters. Thus heart rate variability (HRV) represents one of
the most frequently used parameters, in cardiology. Nowadays there
are different methods of studying heart rate variability dynamics.
In recent years \cite{Guzman}, fluctuations of heartbeat dynamics
have been studied by means of several methods derived from
nonlinear dynamics and statistical physics, such as detrended
fluctuation analysis (DFA) \cite {Peng1, Peng2}, spectral analysis
\cite{Iyengar}, entropy \cite{Richman, Kaplan, Zebrowski},
correlation dimension \cite{Govindan}. In paper \cite{Stanley1}
authors illustrate the problems related to the physiological
signal analysis with representative examples of human heartbeat
dynamics under healthy and pathological conditions which is based
on two methods: power spectrum and detrended fluctuation analysis.
In this paper different characteristics of heartbeat: $1/f$
fluctuations, long-range anticorrelations (monofractal analysis),
self-similar cascades, multifractality and nonlinearity are
considered. By means of a wavelet-based multifractal formalism it
is shown that healthy human heartbeat dynamics exhibits higher
complexity which is characterized by a broad multifractal
spectrum. In paper \cite{McCaffery} multiresolution wavelet
analysis has been used to study the heart rate variability in a
patient with different pathological conditions. Noise effects of
abnormal heartbeats were considered in paper \cite{Stanley2}. The
correlation exceptions of heartbeat dynamics of different sleep
stages often have been researched lately. In papers \cite{Bunde,
Stanley3} correlation properties of the magnitude and the sign of
increments in the time intervals between successive heartbeats
during a light sleep, a deep sleep, a rapid eye movement sleep
were discovered by means of the detrended fluctuation analysis.
Multiscaled randomness \cite{Hausdorff}, multifractal analysis
\cite{Stanley4}, simulation by non-linear oscillators
\cite{Babloyantz}, fractal approach based on scaling of a
frequency spectrum on power law $1/\omega^{\alpha}$
\cite{Stanley5}, quantitative analysis \cite{Kurths} are also used
to analyze heart rate variability. The change of correlations and
statistical memory effects is one of the must important questions
\cite{Allegrini}  in heart rate variability dynamics observed with
ageing \cite{Guzman, Iyengar}.

Among existing methods of researching HRV one can differentiate
the methods of estimating HRV in a time area, spectral methods of
estimating HRV in a frequency area, as well as nonlinear methods.
The last group of methods  has proved to be a powerful means to
study various complex systems and has brought about significant
achievements in processing biological and medical data. In recent
years universal methods of statistical physics have been more
often used  in medicine and biology. The methods of statistical
physics which have been used to research real complex systems
\cite{Yulm1,Yulm2,Yulm3,Yulm4,Yulm5,Yulm6}, in the field of
cardiology reveal essentially new opportunities for the analysis,
diagnostics and forecasting the processes of biological ageing and
diseases of a human heart. They disclose dynamic features of HRV,
latent for classical medical methods of research.

In this paper we offer a new method of study of the problems of
ageing of a human heart activity, based on our theory of discrete
non-Markov processes \cite{Yulm1}. This theory  has already found
practical application in cardiology \cite{Yulm2}, neurophysiology
\cite{Yulm3, Yulm4}, the study of locomotor and sensomotor
activity \cite{Yulm3}, epidemiology \cite{Yulm5} and seismology
\cite{Yulm6}.

\section{Basic concepts and definition of the statistical theory of
nonstationary discrete non-Markov processes  in complex systems}

The obtained data were processed by means of the  above declared
technique. We use the results of our recent theory of discrete
non-Markov random processes for the quantitative description  of
Markovian and non-Markovian components in stochastic alteration of
the registered data. The set of three memory functions was
calculated for each sequence of the data. Frequency power spectra
for each of these functions are obtained by using the  fast
Fourier transform. For a more detailed analysis  of the properties
of the system  we also consider the frequency spectrum of the
first three points of the statistical spectrum of the
non-Markovity parameter. The spectrum of the non-Markovity
parameter was introduced earlier in the following articles
\cite{Yulm1, Yulm2, Yulm6}. In this study we use the frequency
spectrum of the non-Markovity parameter: \bn \varepsilon_i(\omega)
= \left\{ \frac {\mu _ {i-1} (\omega)}{\mu_i (\omega)} \right\}^{
{1}/{2}}, ~~
 \mu_i(\omega)=\left|\int_{0}^{\infty}dt M_i(t)\cos( \omega t)\right|^2=
 \left|\sum_{j=0}^{N-1}M_i(t_j)\cos(\omega t_j)\right|^2,
\nonumber \en here $i=1, 2, 3 ... $ is the number of the
relaxation level, $ \mu_i (\omega)$ is the Fourier-transform and a
power spectrum of the $i$th level memory function $M_i (t)$ (see,
Eq. \Ref{f1} below). The parameters $ \varepsilon_i(\omega) $
allow to receive the quantitative estimation of long-term memory
effects in the experimental time series of the data as shown in
Ref. \cite{Yulm7}. From the physical point of view the parameter $
\varepsilon_i(\omega) $ allows to mark out the three most
important cases \cite{Yulm7}. Markov and completely randomized
processes correspond to values $ \varepsilon \to \infty$,
quasi-Markov processes (memory effects can be noticed there )
correspond to values $ \varepsilon
> 1 $.  The limiting case with $ \varepsilon \sim 1 $ concerns the
situation with non-Markov processes, i.e., processes, where there
is  long-range memory.

In early works \cite{Yulm1, Yulm2, Yulm6} we came to the following
chain of connected non-Markov finite-difference kinetic equations
($t=m\tau $): \be \frac {\Delta M_n (t)} {\Delta t} = \lambda _
{n+1} M_n (t) -\tau
\Lambda _ {n+1} \sum _ {j=0} ^ {m-1} M _ {n+1} (j\tau) M_n (t-j\tau). \\
\label{f1}\ee Here parameters $ \lambda _ {n+1} $ represent eigen
values of the Liouville's quasioperator. The relaxation parameters
of $ \Lambda _ {n+1} $ are determined as follows: \bn
\lambda_{n+1}=i \frac {\langle {\bf W} _n \hat L {\bf W} _n
\rangle} {\langle \left| {\bf W} _n \right| ^ 2 \rangle}, \
\Lambda_n =i \frac {\langle {\bf W} _ {n-1} \hat L {\bf W} _n
\rangle} {\langle \left| {\bf W} _ {n-1} \right| ^2 \rangle}.
\label{f2}\en The zero order memory function $M_0 (t)$ in Eq.
\Ref{f1}: \be M_0 (t) = a (t) = \frac {\langle {\bf A}_k^0 (0)
{\bf A}_{m+k} ^m (t) \rangle} {\langle \left| {\bf A} _k^0 (0)
\right| ^2 \rangle}, ~~ t=m\tau, \nonumber \ee  \be {\bf A}_k^0
(0) = (\delta x_0, \delta x_1, \delta x_2, \ldots, \delta x _
{k-1}), \nonumber \ee \be {\bf A}_ {m+k} ^m (t) = \{\delta x_m ,
\delta x _ {m+1}, \delta x_{m+2}, \ldots, \delta x _ {m+k-1}\},
\nonumber \ee describes statistical memory in complex systems with
a discrete time (${\bf A}_k^0 (0)$ and ${\bf A}_ {m+k} ^m (t)$ are
vectors of the initial and final states of the studied system). In
paper \cite{Yulm1}  we have received the recurrent formula on the
basis of Gram-Schmidt orthogonalization procedure, in which the
senior dynamic variable $ {\bf W} _n = {\bf W} _n (t) $ is
connected with the junior one in the following way: \bn { \bf W}
_0 = {\bf A}_k^0
(0), ~~{\bf W}_1 = \{ i \hat L-\lambda_1 \}{\bf W}_0, \ldots ~~ \\
\nonumber {\bf W}_n = \{ i\hat L-\lambda _ {n-1} \} {\bf W}_{n-1}+
\Lambda_{n-1} {\bf W}_ {n-2}+..., ~~ n > 1. \label {f3} \en

The initial time correlation function (TCF) $a(t)$ and the set of
discrete memory functions $M_n (t) $ in Eq. \Ref{f1} are important
for further consideration. The first three equations of this chain
($t=m\tau $ is a discrete time) can be presented as follows: \bn
\frac {\Delta a (t)} {\Delta t}&=-&\tau \Lambda_1 \sum _ {j=0} ^
{m-1}
M_1 (j\tau) a (t-j\tau) + \lambda_1 a (t), \nonumber \\
\frac {\Delta M_1 (t)} {\Delta t}&=-&\tau \Lambda_2 \sum _ {j=0} ^
{m-1}
M_2 (j\tau) M_1 (t-j\tau) + \lambda_2 M_1 (t), \nonumber \\
\frac {\Delta M_2 (t)} {\Delta t}&=-&\tau \Lambda_3 \sum _ {j=0} ^
{m-1} M_3 (j\tau) M_2 (t-j\tau) + \lambda_3 M_2 (t). \label{f4}\en
This system of finite-difference Eqs. \Ref {f1}, \Ref{f4} is a
discrete analogue of the well-known chain of kinetic
Zwanzig'-Mori's equations. The latter plays the fundamental role
in modern statistical physics of non-equilibrium phenomena with a
continuous time. It is necessary to note that the chain of
Zwanzig'-Mori's equations  is valid only for quantum and classical
Hamiltonian systems with a continuous time. The chain of
finite-difference kinetic Eqs. \Ref {f1}, \Ref{f4} is valid for
complex systems, in which there is no Hamiltonian, the time is
discrete, and there are no exact equations of motion. However,
"dynamics" and "motion" in real complex systems undoubtedly exist
and can be registered in the experiment. The first three equations
in the chain \Ref {f4} form a basis for the quasihydrodynamic
description of stochastic discrete processes in complex systems.
The application of Eq. \Ref{f1} opens up new possibilities in the
detailed analysis of the statistical properties of correlations in
complex systems. The  existence of finite-difference Eqs. \Ref
{f1}, \Ref{f4} allows to evaluate unknown memory functions
(similarly time correlation functions) directly from the
experimental data.

Let's determine the experimental relaxation time $\tau_E $ by the
equation: \be \tau_E=\Delta t \sum_{j=1}^{N} a(t_j). \ee
\\
Using the experimental data now we can define the relaxation time
of the studied system. Further we can compare the experimental
time $\tau_E $ with the theoretical one. The theoretical
relaxation time $\tau_E $ (where $i=1,2,3...$ is the number of
approximation) can be determined on the basis of Zwanzig'-Mori's
equations for various correlation approximations. For the first
age group (young people) better accordance between the
experimental and theoretical times is gained in the 6th
correlation approximation: $\ M_3 (t) = M_1 (t)$ (see Table 1).
This one shows the presence of long-range memory in the considered
system. For the examined group of elderly persons better
accordance of relaxation times is received in the first
correlation approximation $\ M_1(t) = a(t)$ (see Table 1). It
indicates to the existence of short-range memory in this group.

Using the Laplace transform on the  first three Zwanzig'-Mori's
equations,  we shall receive: \be \nonumber
s\tilde{a}(s)-1=\lambda_{1}\tilde{a}(s)-\Lambda_{1}\tilde{a}(s)\tilde{M_{1}}(s),
\\ \nonumber \ee
\be
s\tilde{M_{1}}(s)-1=\lambda_{2}\tilde{M_{1}}(s)-\Lambda_{2}\tilde{M_{1}}(s)\tilde{M_{2}}(s),
\\ \nonumber \ee
\be
s\tilde{M_{2}}(s)-1=\lambda_{3}\tilde{M_{2}}(s)-\Lambda_{3}\tilde{M_{2}}(s)\tilde{M_{3}}(s).
\\ \nonumber \ee
One can solve this system by means of various approximations. For
the approximation $ M_{1}(t)=a(t)$ (the first approximation in
Table 1) we shall receive:

\be
\\ \nonumber
\tilde{a}(s)=\frac{-(s-\lambda_1)+\sqrt{(s-\lambda_1)^2+4\Lambda_1^2}}{2\Lambda_1},
\\ \nonumber \ee
\be
\\ \nonumber
\tau_1=\lim_{s \to
0}\tilde{a}(s)=\frac{\lambda_1+\sqrt{\lambda_1^2+4\Lambda_1^2}}{2\Lambda_1}.
\\ \nonumber \ee
\bigskip \scriptsize
\begin{flushleft}
\footnotesize Table 1
\\
The experimental $ \tau_E $ and theoretical $ \tau_i$ ($i=1,\ldots,6$)
relaxation times for different age groups
\end{flushleft}
\begin{tabular}{p{1cm}p{1cm}p{2.4cm}p{2.4cm}p{2.4cm}p{2.4cm}p{2.4cm}p{2.4cm}}
\hline Age&$\tau_E$&$\tau_{1}$(${\tiny {M_{1}(t)=}}$&$\tau_{2}{\scriptsize
(M_{2}(t)=}$&$\tau_{3}{\scriptsize
(M_{3}(t)=}$&$\tau_{4}{\scriptsize
(M_{2}(t)=}$&$\tau_{5}{\scriptsize (M_{3}(t)=}$
&$\tau_{6}{\scriptsize (M_{3}(t)=}$
\\
&&$\tiny{a(t)}$)&$\tiny{M_{1}(t)}$)&$\tiny{M_{2}(t)}$)&
$\tiny{a(t)}$)&$\tiny{a(t)}$)
&$\tiny{M_{1}(t)}$)
\\ \hline
 Young&0.126&0.987&1.171&1.172&1.188&1.219&0.211
\\
 Old&0.147&0.260&7.262&6.864&0.789&5.440&0.045
\\ \hline
\end{tabular}
\normalsize
\\

For the approximation $ M_{3}(t)=M_{1}(t)$ (the sixth
approximation in Table 1) we shall find: \be
\\ \nonumber
\tilde{a}(s)=\frac{1}{\Lambda_{1}\tilde{M_{1}}(s)+s-\lambda_{1}}=
\\ \nonumber \ee
\footnotesize \be
\frac{2\Lambda_{3}(s-\lambda_{2})}{2\Lambda_{3}(s-\lambda_{2})(s-\lambda_{1})-\Lambda_{1}(\Lambda_{2}+(s-\lambda_{2})(s-\lambda_{3})-\Lambda_{3})+\Lambda_{1}\sqrt{(\Lambda_{2}+(s-\lambda_{2})(s-\lambda_{3})-\Lambda_{3})^{2}+4\Lambda_{3}(s-\lambda_{2})(s-\lambda_{3})}},
\nonumber \ee \normalsize \be \tau_{6}=\lim_{s \to
0}\tilde{a}(s)=\frac{2\lambda_{2}\Lambda_{3}}{\Lambda_{1}(\Lambda_{2}+\lambda_{2}\lambda_{3}-\Lambda_{3})-2\lambda_{1}\lambda_{2}\Lambda_{3}-\Lambda_{1}\sqrt{(\Lambda_{2}+\lambda_{2}\lambda_{3}-\Lambda_{3})^{2}+4\lambda_{2}\lambda_{3}\Lambda_{3}}}.
 \nonumber
\ee

In Table 1 we show the experimental times of relaxation $ \tau_E $
(see Eq. (5)) and theoretical times of relaxation $\tau_i$ for
different age groups.

\section{Experimental data}

We used the time series of R-R intervals in young and elderly
subjects as the experimental data \cite{Iyengar}. Two groups of
healthy human subjects: 10 young (mean age 27 yr, range 21-34 yr)
and 10 elderly (mean age 74 yr, range 68-81 yr), participated in
this study. Each group consisted of five women and five men. All
subjects provided written informed consent and underwent a
screening history, physical examination, routine blood count and
biochemical tests, electrocardiogram, and exercise tolerance test.
Only healthy, nonsmoking subjects with normal exercise tolerance
tests, without any medical problems, and being on no medication
were admitted to the study.

All subjects remained in an inactive state in sinus rhythm while
watching the movie "Fantasia" (Disney) to maintain wakefulness.
Each heartbeat was annotated by means of an automated arrhythmia
detection algorithm, and each beat annotation was verified by
visual inspection. The R-R interval (interbeat interval) of  time
series for each subject was then computed \cite{Iyengar}.

\section{Discussion of the results}
The basic outcomes are submitted in this section. Further the
appropriate analysis of the experimental data will be carried out
both for young and old people. Two new qualitative procedures of
the appropriate analysis have been used. The procedure of the
window-time behavior shows great oscillations of an R-R interval
for the power spectra of memory functions with respiratory
arrhythmia. The calculation of the local relaxation parameters is
carried out by means of the procedure of time localization. The
decrease of the relaxation rate in ageing people is indicated
through the time dependence of the localized relaxation
parameters. The quantitative assessment of non-Markovity effects
of heart rate variability is carried out  by means of a special
measure.

\subsection{Study of age-related alterations of heart rate variability}
Further we submitted figures for one young and one old person. The
figures reflect a general pattern of the group. Below we show the
analysis which describes the experimental data for the first and
second group.

\begin{figure}\includegraphics[width=12cm, height=8cm,angle=0]{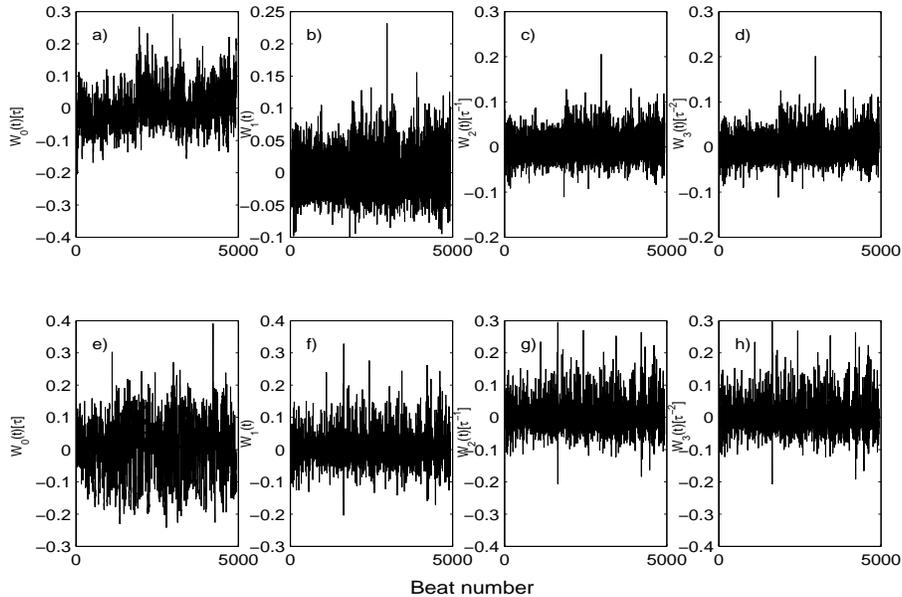}
\caption{The time series of the first four  dynamic orthogonal
variables $W_i $, where $i=0,..,3 $ for the young (a-d) and the
elderly persons (e-h). The time series of the elderly person is
accompanied by more significant fluctuations of an R-R interval.
The time series of elderly person is characterized by greater
frequency of occurrence of significant fluctuations. It shows
higher degree of Markovity of the fluctuations of an R-R interval
for the elderly person.}
\end{figure}

In Fig. 1 we have presented the time series of  the first four
dynamic orthogonal variables $W_i$, where $i=0,...,3$ for young
(Figs. 1a-d) and elderly (Figs. 1e-h) people. The time series of
the elderly people differ in greater amplitude and frequency of
fluctuations of an R-R interval. The scattering interval of the
oscillations of an R-R interval for young people constitutes
$(0.83\tau\div1.18\tau$, where $\tau=1.0244 s$, $\tau$-time of
discretization). The variability of an R-R interval in elderly
people changes within the limit of $(0.92\tau\div1.5\tau$, where
$\tau=1.062s)$ for the initial signal. The frequency of
fluctuations of an R-R interval in elderly people is greater than
in young people. The normal cardiac activity of elderly people is
accompanied by more randomized fluctuations of an R-R interval.

In Fig. 2 we have presented the phase portraits for the
variability of an R-R interval in young and elderly people on six
plane projections of the first four  dynamic orthogonal variables.
The phase clouds of the young and elderly people are symmetric
concerning the origin of coordinates frame. The phase clouds have
a centered nucleus. The nucleus encloses some points, speckled on
the perimeter. The phase points for elderly people have a 2-3
times greater interval of a scattering,  due to the presence of
more appreciable oscillations of an R-R interval.

\begin{figure}\includegraphics[width=12cm, height=9cm,angle=0]{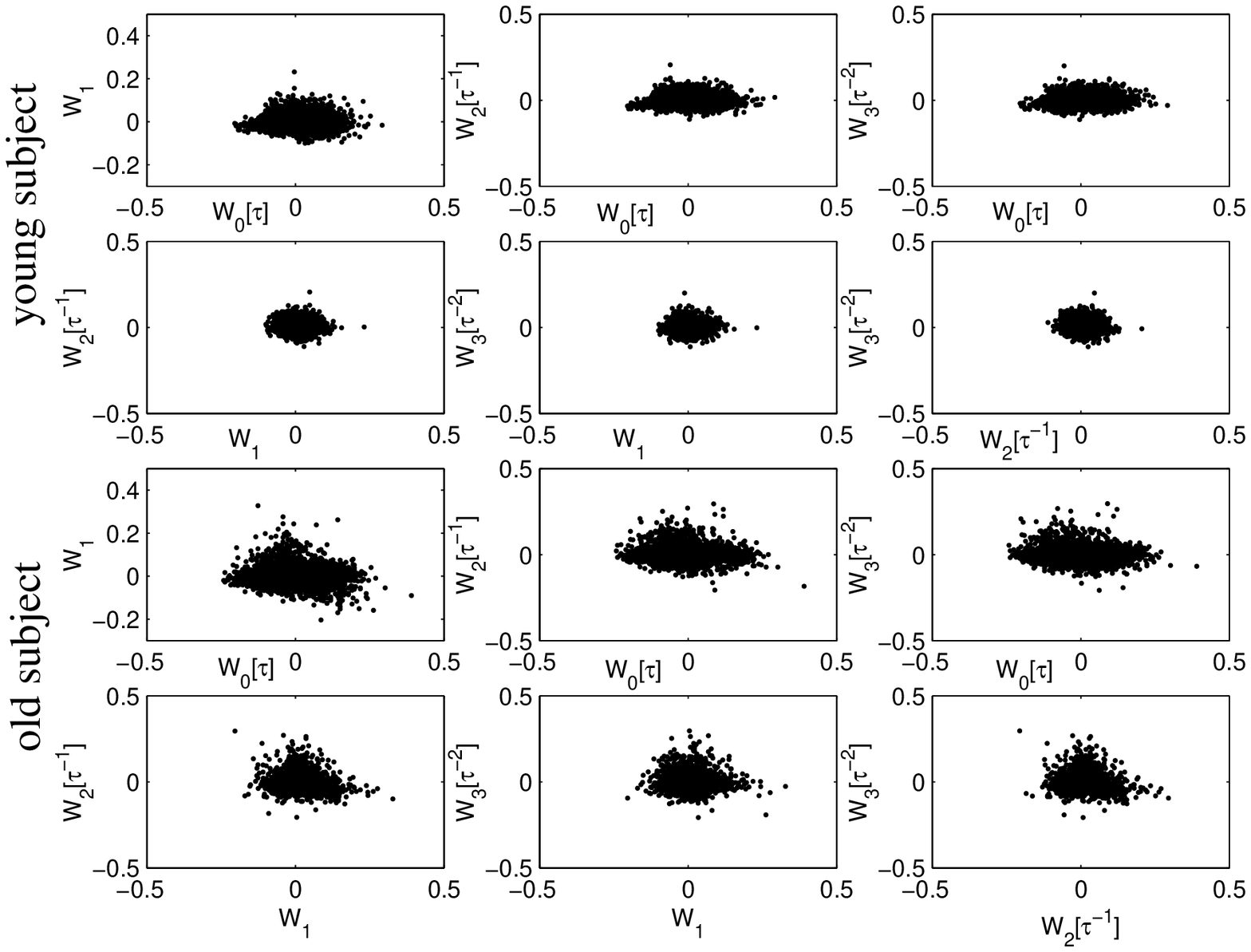}
\caption{The phase portraits of variability of an R-R interval on
six plane projections of various combinations $W_i, W_j $ for
young and elderly people. The phase clouds of the young man are
more compressed towards the center of coordinates system. Around
the centralized nucleus of the phase clouds for the elderly
person, separate points are scattered. The interval of their
disorder 2-3 times exceeds the areas of disorder of the phase
points for the young man.}
\end{figure}

\begin{figure}\includegraphics[width=12cm, height=8cm,angle=0]{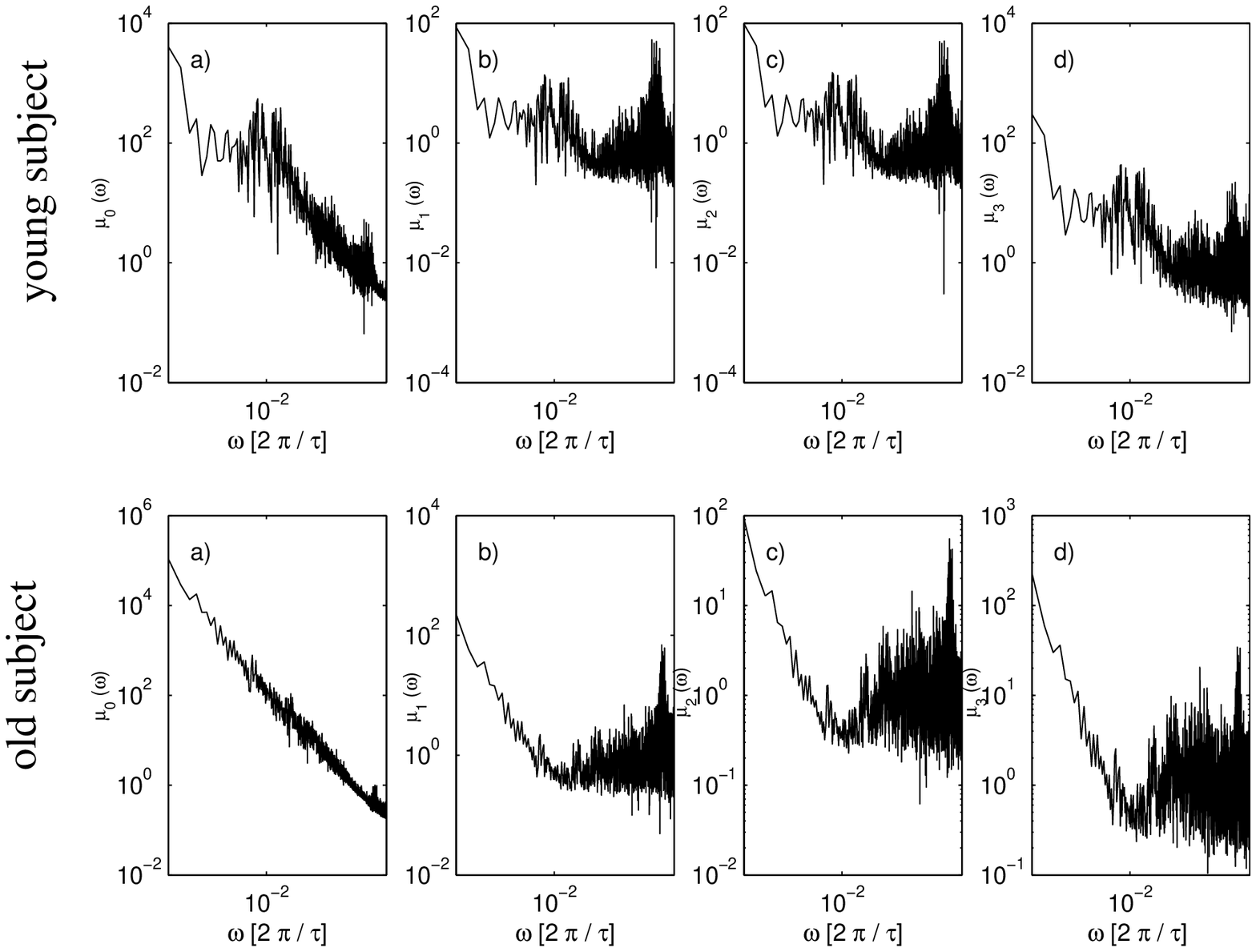}
\caption{The power spectra of the initial TCF (a) and  the first
three memory functions (b-d) for the young and elderly persons.
The power spectra of the initial TCF of the elderly person differs
in smaller dynamic breaks in the field of low frequencies. In all
the power spectra the dynamic splashes (peaks) are found in the
field of high frequencies, which is connected with respiratory
arrhythmia. These dynamic peaks of the elderly person are shifted
in the area of higher frequencies. It testifies to the increase in
frequency of cardiac reductions at breath that comes with ageing.}
\end{figure}

In Fig. 3 we have presented the power spectra of the initial TCF
$\mu_0(\omega)$ (Fig. 3a) and the first three memory functions of
younger orders $\mu_i(\omega)$ (where $i=1,2,3$) (Figs. 3b-d) for
young   and elderly people. The figures are submitted on a log-log
scale. The power spectrum of the initial TCF for the elderly
person differs in smaller dynamic fractures in the area of low
frequencies. It is possible to find dynamic splashes (dynamic
peaks) on all figures in the area of frequencies $(0.2
f.u.<\omega<0.5 f.u.$, where $1 f.u.=1/\tau$), in particular.
These dynamic peaks appear due to respiratory arrhythmia. The
given dynamic splashes remind of the well known shape of the
Suyumbike Tower \cite{Yulm2}. The increase of the power spectrum
on these frequencies reflects age-physiological changes. Ageing
people develop a shift of these dynamic peaks in the range of high
frequencies. The dynamic peaks, which relate to the respiratory
arrhythmia of the young man, are discovered within the frequency
interval of $0.25 f.u.<\omega<0.45 f.u.$ These dynamic splashes of
the elderly person are in the range of $0.4 f.u.<\omega<0.55
f.u.$. The specific arrhythmia of cardiac activity at respiration
accounts for this conclusion. This frequency (at respiratory
arrhythmia) in elderly persons is higher than in young ones.

The procedure construction of the window-time behavior of the
power spectra of memory functions leads to the similar conclusion.
The similar procedure allows to consider in detail any dynamic
regularities originating in the power spectra of the memory
functions. The idea of this procedure consists in the following
\cite{Yulm4}. Originally it is necessary to determine the optimal
length of the sample. When the length of the sample is small, the
"accumulated" information will be insufficient for carrying out a
qualitative correlation analysis due to gross errors and the
influence of noise effects. When the length of the sampling is
large the necessary "sensitivity" weakens. The analysis of samples
of different lengths shows, that the optimal length for this
procedure constitutes $2^8=256 $ points. From the initial array of
the experimental data we take 256 initial points. We receive the
first window of 256 points. Then we build a memory function power
spectrum for this sample and take the next window of 256 points
(from 257 to 512). Then we build the power spectrum of the memory
function. This procedure is carried out repeatedly up to the end
of the sampling of the experimental data. In Fig. 4 the
time-window behavior of the first memory function $ \mu_1 (\omega)
$ in young and elderly people is submitted. The most appreciable
dynamic splashes (peaks) in the power spectra are connected with
respiratory arrhythmia. All dynamic peaks are found in the
particular range of frequencies. Generally in young people the
range of these dynamic peaks meets $0.25 f.u. <\omega <0.45 f.u.$.
For elderly people this range is shifted  to the right and meets
$0.4 f.u. <\omega <0.55 f.u.$ This implies amplification of
cardiac activity at respiration with ageing.

\begin{figure}\includegraphics[width=12cm, height=8cm,angle=0]{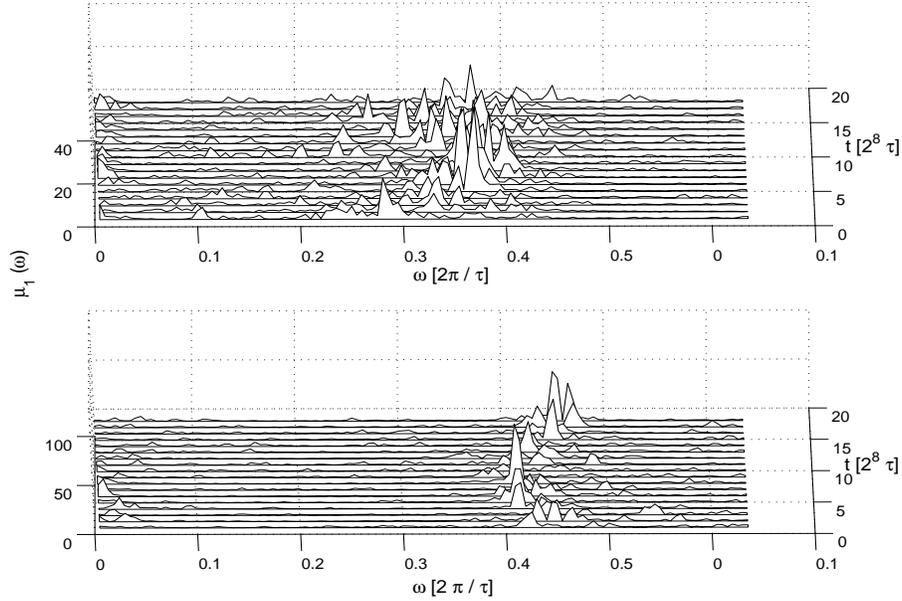}
\caption{The window-time behavior of the first memory function
power spectra $ \mu_1 (\omega) $ for young and elderly people. On
certain frequencies the dynamic peaks are distinctly visible in
both power spectra. These peaks are connected with respiratory
arrhythmia. These peaks are shifted towards higher frequencies for
elderly people. It confirms the conclusion made as a result of the
analysis of the previous figure.}
\end{figure}

\subsection{Quantitative measure of the effects of non-Markovity
in heart rate variability}

Any complex system has a great number of degrees of freedom. The
high dimension of complex systems, the presence of strong
nonlinear interactions and the feedback determine their behavior.
This behavior can be characterized by Markov random processes.
Strong external influence at accidents, crises and human diseases
entails partial synchronization of natural chaotic behavior of
complex systems. This synchronization results in the compelled
organization of the structure of a real system and the occurrence
of regular communications. The behavior of the system becomes more
ordered. Such behavior is defined by the amplification of
non-Markov statistical effects.

\begin{figure}\includegraphics[width=12cm, height=8cm,angle=0]{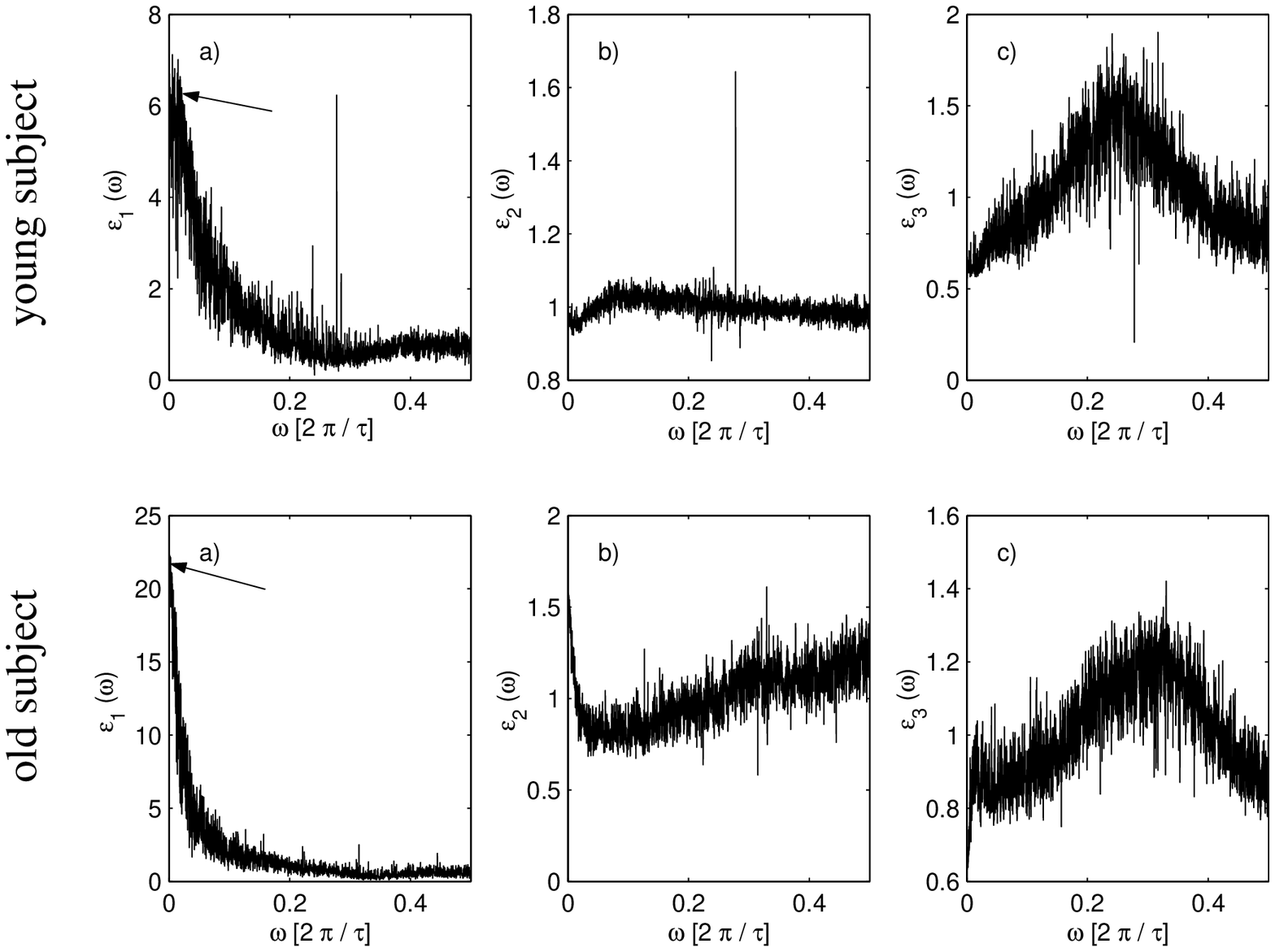}
\caption{The frequency dependence of first three points of the
non-Markovity parameter $ \varepsilon_i (\omega) $ for the young
and elderly persons. The value of the first point of the
non-Markovity parameter $ \varepsilon_1 (0) $ (a) on zero
frequency is an original quantitative measure of non-Markovity of
the process under study. The value of this parameter for the young
man is equal to 6.24. For the elderly person the value of this
parameter is equal to 23.16. The comparison of these values
indicates, that the heart activity of the elderly person is
characterized by the greater Markovity.}
\end{figure}

The basic idea of our method consists in defining the quantitative
proportion between Markov and non-Markov effects of the studied
stochastic process. As a quantitative measure of non-Markovity we
suggest using the first point of the non-Markovity parameter  $
\varepsilon_1 (\omega) $, where $ \omega=0 f.u. $ The physical
sense of this parameter consists in distinguishing Markov
(processes with instant or short memory) and non-Markov (processes
with long-range memory) stochastic processes.  The increase of
this parameter ($ \varepsilon_1 (0)>> 1 $) means greater Markovity
of conditions of the system, the reduction of this parameter
describes the processes with statistical effects of non-Markovity.
The increase or decrease of the non-Markovity parameter makes it
possible to judge the degree of non-Markovity in the complex
system. Thus non-Markov effects are characterized by the presence
of long-range correlations and the amplification of statistical
displays of long-range memory. Markovian processes are
characterized by the amplification of short-range correlations and
the reduction of the effects of statistical memory. The
quantitative measure of non-Markovity $\varepsilon_1 (0) = \left\{
\frac {\mu _ 0 (0)} { \mu_1 (0)} \right\}^{\frac {1} {2}}$ reveals
the nature of the behavior of a system.

In Fig. 5 the frequency dependence of the first three points of
the non-Markovity parameter $ \varepsilon_i (\omega) $, where
$i=1,2,3 $ in young and elderly people, is submitted. The value of
a quantitative measure of the degree of non-Markovity of the young
man is $ \varepsilon_1 (0) =6.24 $. The value of this parameter of
the elderly person is $ \varepsilon_1 (0) =23.16 $. The ratio of
the quantitative measures of non-Markovity is 3.7 times. It
testifies to the increase of Markovian effects in fluctuations of
an R-R interval with ageing. Markovity of heart rate variability
is connected with significant fluctuations in the initial signal
of the elderly person.

Further we have presented statistical results of processing for
the first and second age groups. In Fig. 6 we have presented the
first points of the non-Markovian parameter, averaged for the
groups for ten young and ten elderly people. The frequency
dependence of such parameters is defined as follows:
 \bn \varepsilon_i(\omega)_{av.val}=\frac
{\sum \limits_{j=1}^{N}\varepsilon_{i,j}(\omega)}{N}, i=1,...,3,
\nonumber
\\ \varepsilon_1(\omega)_{av.val}=\frac {\sum
\limits_{j=1}^{10}\varepsilon_{1,j}(\omega)}{10}, ...\nonumber
\en The value of the first point of the non-Markovity parameter  $
\varepsilon_1 (0) _ {av.val} $ of young people is 6.41.  The value
of this parameter  for elderly people is equal to 27.41. The ratio
of these values is 4.2 times. It testifies to the amplification of
Markov effects in the heart activity in elderly people. Therefore
the number of  Markov components affecting the heart activity
increases with ageing.

\begin{figure}\includegraphics[width=12cm, height=8cm,angle=0]{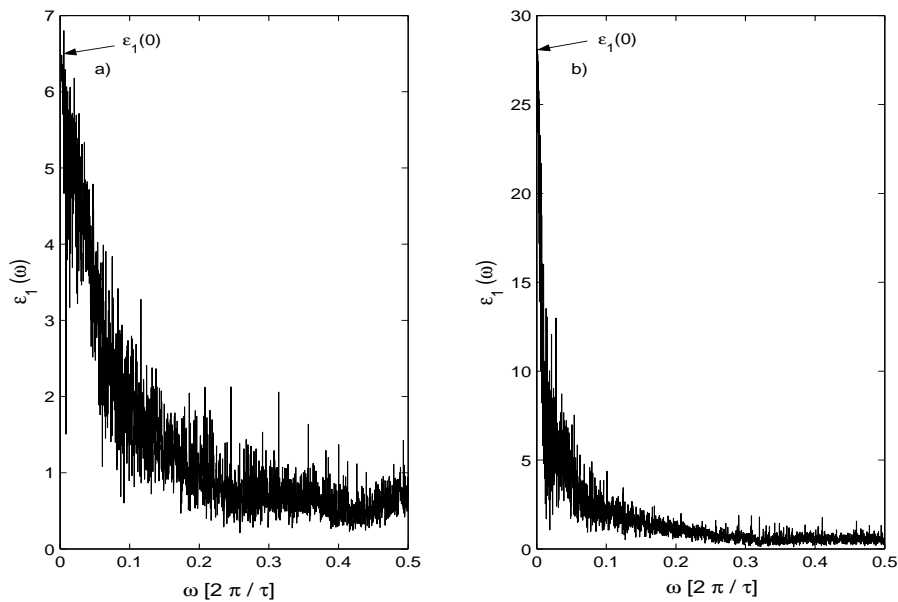}
\caption{The frequency dependence of the first point of
non-Markovity parameter, averaged on the group of young (a) and
elderly (b) people. These figures allow to define the generalized
degree of Markovity and non-Markovity for the first and second age
groups. The ratio of quantitative measures of the degree of
non-Markovity $ \delta =\frac {\varepsilon_1 (0) _ {old}}
{\varepsilon_1 (0) _ {young}} $ for the second and first age
groups constitutes 4.2 times. This implies that the variability of
an R-R interval of elderly people becomes more Markovian (for the
whole group). The heart activity of young people becomes more
ordered and is characterized by high regularity.}
\end{figure}

\subsection{Age-related alterations of relaxation modes}
The local relaxation parameters allow to estimate the relaxation
rate in the systems. The procedure of localization enables to
reveal internal peculiarities of the dynamics of cardiac activity,
latent for usual correlation analysis. The idea of the method
consist in the following. From the initial time series we take a
sampling N points in length for which we calculate kinetic and
relaxation parameters. Then we performed  "step-by-step shift to
the right" operation one interval to the right and calculate
kinetic and relaxation parameters. This procedure is carried out
up to the end of the time series. Thus received local relaxation
parameters have high sensitivity effects of alternation and
non-stationarity. If there is any irregularity in the initial time
series it will be instantly revealed in the time behavior of local
parameters.

\begin{figure}\includegraphics[width=12cm, height=8cm,angle=0]{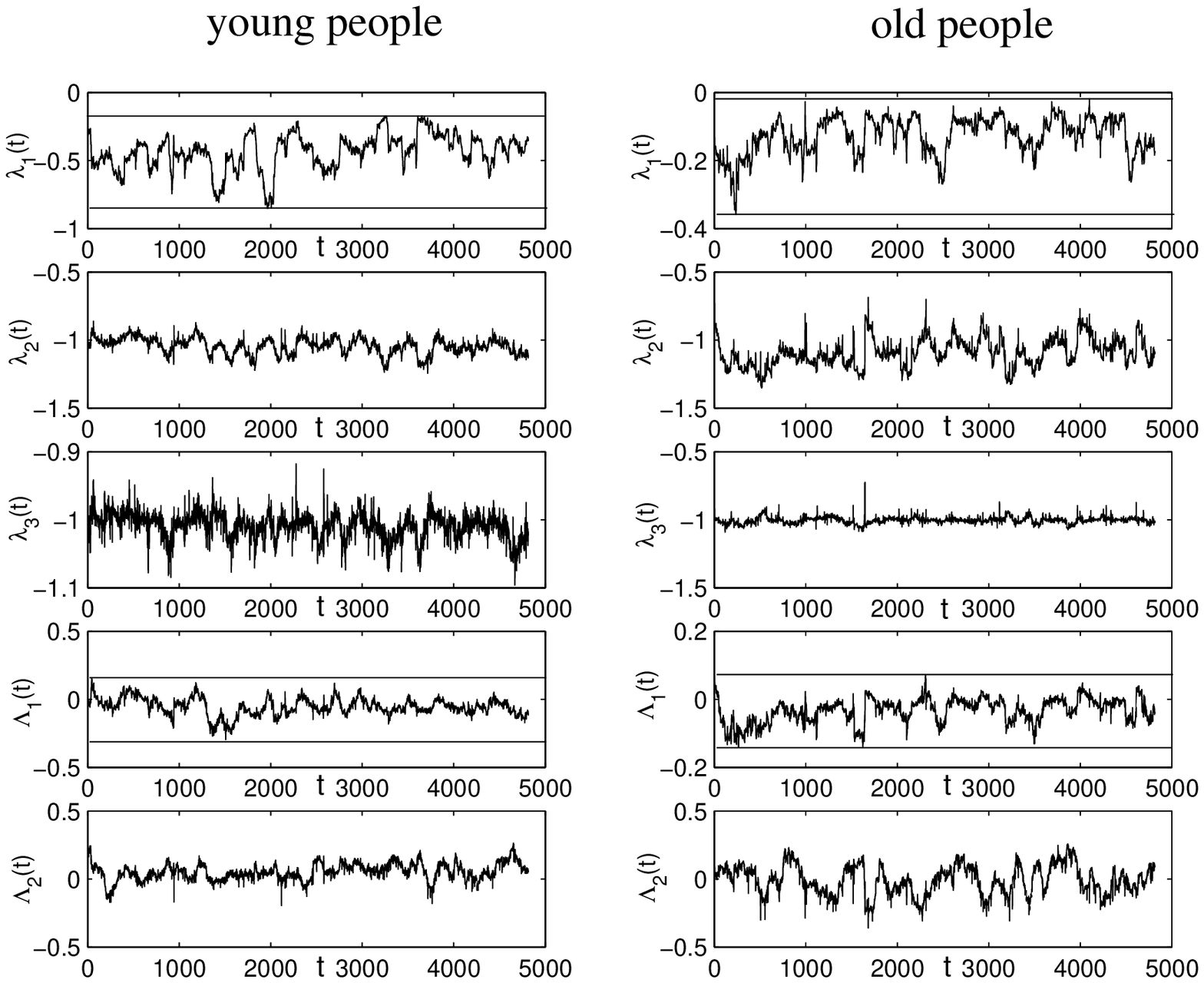}
\caption{The time dependencies of local kinetic and relaxation
parameters $ \lambda_1, \lambda_2, \lambda_3 $ and $ \Lambda_1,
\Lambda_2 $, averaged on the whole group for young and elderly
people. The physical sense of the first relaxation parameter $
\lambda_1 $ consists in defining the relaxation rate. The ratio of
root-mean-square amplitude of this parameter to the first and
second age groups is equal to 3.3 times. It indicates to higher
relaxation rate of cardiac activity for the young people.}
\end{figure}

When using this method (as well as in the first procedure) it is
necessary to define the length of the sample which allows to
receive the most  trustworthy information. As a result of the
research of different lengths of the local sample we have
calculated the optimal length which contains $2^7=128 $ points.

In Fig. 7 we have presented the time dependence of local kinetic
and relaxation parameters $\lambda_i$, where $i=1,2,3$ and
$\Lambda_1, \Lambda_2$  averaged for ten young and ten elderly
people. The physical sense of the first relaxation parameter
$\lambda_1$ consists in defining the relaxation rate for the
studied process. On average the amplitude of local parameter
$\lambda_1$ for young people changes within the interval of
$0.1719 \tau^{-1}<\lambda_1<0.8522 \tau^{-1}$. For elderly people
this variable changes within the interval of $0.0187
\tau^{-1}<\lambda_1<0.3580 \tau^{-1}$. The ratio of
root-mean-square amplitude $\langle A \rangle=\left\{\frac {\sum
\limits_{j=0}^{N-1} x_j^2}{N}\right\}^{\frac {1} {2}}$ for young
and elderly people is equal to 3.3  times. In Table 2 we present
the root-mean-square amplitude, dispersion
$\sigma^2=\frac{1}{N}{\sum \limits_{j=0}^{N-1} \delta x_j^2}$ and
a root-mean-square deviation $\sigma=\left\{\frac{1}{N}{\sum
\limits_{j=0}^{N-1}(x_j-\langle x
\rangle)^2}\right\}^{\frac{1}{2}}$ for local relaxation
parameters. The comparison of these characteristics indicates
reduction of the relaxation rate with ageing (the increase of the
relaxation time). For example, the difference of the amplitudes of
relaxation parameters $ \Lambda_1 $ constitutes 1.7 times.

\begin{flushleft}
\footnotesize Table 2
\\ Some kinetic and relaxation parameters
(absolute values) for young and elderly people, calculated from
our theory
\end{flushleft}

\begin{center}
\begin{tabular}{p{2cm}p{1.3cm}p{1.3cm}p{1.3cm}p{1.3cm}p{1.3cm}p{1.3cm}p{1.3cm}p{1.3cm}p{1.3cm}p{1.3cm}}
\hline Age & young&old& young & old & young & old &
young&old&young&old
\\ \hline
 Parameter  & $\lambda_1(\tau^{-1})$& & $\lambda_2(\tau^{-1})$& & $\lambda_3(\tau^{-1})$& & $\Lambda_1(\tau^{-2})$ && $\Lambda_2(\tau^{-2})$& \\
\hline$\langle A \rangle$ & 0.4551& 0.1384& 0.0490&1.1021 &1.009
&1.004&0.083&0.048&0.083&0.107
\\
$\sigma^2$ &0.016&0.003&0.004&0.010&$3*10^{-4}$&$5*10^{-4}$&0.004&0.001&0.004&0.011 \\

$\sigma$&0.128&0.054&0.060&0.101&0.018&0.024&0.063&0.036&0.062&0.107\\
\hline
\end{tabular}
\end{center}

Thus, using different approaches in the research of the time
series of an R-R interval we have arrived at the general
conclusion. The work of a heart becomes more Markov (the effects
of statistical memory disappear) with ageing and the speed of
relaxation of cardiac activity is reduced.
\section{Conclusion}
The achieved results allow to come to the following conclusions.
The procedure of localization makes it possible to calculate
quantitative characteristics describing the speed of relaxation of
cardiac activity. We have revealed essential distinctions in
relaxation processes for different age groups on the basis of the
comparative analysis of quantitative characteristics of
variability of an R-R interval. The processes occurring in the
heart of a young person, have a greater relaxation rate. Hence, at
the appearance of any destructions in cardiac activity, faster
restoration of its usual normal mode can be observed. The
relaxation rate of cardiac activity decreases in elderly people.
Heart activity comes back to its normal rhythm slower in this
case.

The use of the first point of the non-Markovity parameter $
\varepsilon_1 (\omega) $ allows to estimate quantitatively
Markovian and non-Markovian effects of heart rate variability. The
work of a heart of an older person is characterized by greater
Markovity due to the influence of a greater number of components,
that reduce the effects of long-range memory (deterioration of an
organism, ageing and physiological changes of a heart and other
life-support systems etc.). The comparison of values of the
parameter $ \varepsilon_1 (\omega) $ for young and elderly people
(their ratio is equal to 4.2 in case of $\omega=0 f.u.$),
indicates a high degree of Markovity of heart rate variability of
elderly people. The heart rate variability of young people is
characterized by greater non-Markovity, that indicates the smaller
number of its Markov components. Thus, the number of Markov
components of cardiac activity increases with age.

These conclusions are interconnected and supplement each other.
The increase of the number of Markov components, affecting cardiac
activity, results in the increase of the relaxation time of a
system. The system needs greater time for restoration to the
normal operating mode. It is caused by the decrease of long-range
correlations and reduction of the effects of statistical memory
with ageing. On the contrary, the increase of the relaxation rate
of a system testifies to the increase of the  number of  regular
components. For example, the effects of statistical memory and
long-range correlations are amplified the dynamics of HRV. A
higher relaxation rate is characteristic of normal heart activity.

The procedure of the window-time behavior allows to find out
additional age features of cardiac activity of a person. The
frequency of cardiac reductions at breath increases with age. It
shows the age-related displacement of dynamic bursts (connected
with respiratory arrhythmia) in the area of higher frequencies.

\section{Acknowledgements}
This work  supported by the RHSF (Grant No. 03-06-00218a), RFBR
(Grant No. 02-02-16146, 03-02-96250) and CCBR of Ministry of
Education RF (Grant No. E 02-3.1-538). The authors acknowledge Dr.
Ary L. Goldberger, Dr. C.-K. Peng for stimulating criticism and
valuable discussion and Dr. L.O. Svirina for technical assistance.

\end{document}